# BYOD Security: A New Business Challenge

Kathleen Downer
School of Computing & Mathematics
Charles Sturt University
NSW, Australia-2640
kathleendowner@gmail.com.au

Maumita Bhattacharya
School of Computing & Mathematics
Charles Sturt University
NSW, Australia-2640
mbhattacharya@csu.edu.au

*Abstract*— Bring Your Own Device (BYOD) is a rapidly growing trend in businesses concerned with information technology. BYOD presents a unique list of security concerns for businesses implementing BYOD policies. Recent publications indicate a definite awareness of risks involved in incorporating BYOD into business, however it is still an underrated issue compared to other IT security concerns. This paper focuses on two key BYOD security issues: security challenges and available frameworks. A taxonomy specifically classifying BYOD security challenges is introduced alongside comprehensive frameworks and solutions which are also analysed to gauge their limitations.

*Keywords—BYOD, BYOD security, BYOD security framework*

## I. INTRODUCTION AND BACKGROUND

BYOD is a relatively new initiative adopted by modern businesses which allows employees to use personal mobile devices to complete work in a convenient and flexible manner. Recent industry reports claim approximately 70% of businesses already utilize BYOD and agree they experience improvements including enhanced productivity, efficiency, morale and reduced hardware expenses [15][35][39]. Of these, 50% of employees actively use pre-installed security measures on their device (eg. pass codes), yet less than 20% utilize extra methods (eg. anti-malware) [5][26][27]. In contrast, the rate of threats and attacks aimed towards mobile devices are increasing; especially software based attacks [19][29][39]. This paper was inspired by inconsistencies in research specifically concerning BYOD security. Analysis of reviewed literature revealed that researchers direct their focus towards security challenges and frameworks which counteract certain threats (see Tables 1, 2 and 3). This information was collated to provide a well-rounded view of the current state of BYOD security. This paper introduces a new taxonomy for categorising BYOD security challenges inspired by those used for classifying network security threats taught by Hansman [18]. The BYOD security challenge taxonomy is divided into two dimensions:

**Dimension 1.** Security challenges are classified according to areas and resources of the organisation they affect most. There are two categories: Equipment (software and hardware) based and Human resource challenges.

**Dimension 2.** Further divides challenges by primary concerns, key characteristics, similarities and logical relationships. Equipment based challenges are further divided into 'deployment challenges' and 'technical challenges'. Deployment challenges occur during pre-implementation, whereas Technical challenges are ongoing concerns throughout the lifecycle of a BYOD strategy. Human resource challenges is divided into 'Policy and regulation challenges' (laws and privacy rights) and 'Human aspect challenges' (issues directly concerning employees).

The paper is organised as follows: Section II categorises BYOD security challenges using the above taxonomy, Section III explores existing frameworks and Section IV exposes their limitations.

## II. BYOD SECURITY CHALLENGES

### A. Deployment Challenge

Determining exactly where and how BYOD is necessary is an initial challenge for companies when implementing security policies [2][7] (see Figure 1).This involves analysing all departments and employee responsibilities, then deciding which resources are accessible by mobile devices. Difficulty arises when determining how data is accessed and controlled when employees job share or when an employee's job encompasses many roles. Mobile devices involved in job sharing are prone to data duplication, as employees may modify data differently.

### B. Technical Challenge

**Access control** for mobile devices coincides with the previous challenge. Companies need to determine permission levels for each employee when accessing certain company resources with personal devices and external network connections [2][7].Other factors that determine access control specifications include: setting time limits, limiting how many people can access certain resources at one time and how employees will gain access to company resources. Access control issues and considerations vary according to the business size, location, number of employees and industry.

**Incorporating security measures to cover a range of portable devices** against threats and attacks is complicated, as employees will own an unpredictable range of devices with differing operating systems, meaning the security needs of each need to be equally supported where possible. Clashes between operating system such as requirements, behaviours, conditions and default security issues, will determine security measures required [8][7]. Constantly adjusting security measures to protect all devices is a heavy strain on resources and personnel responsible for maintaining them.

**Table 1.** Some key publications with the most influence towards BYOD research.

| Research | Focus | Limitations | Survey/Review (S/R) / Investigation (I) |
|---|---|---|---|
| Bradford Networks, 2012 | Explains security challenges and guidelines for forming BYOD policies. | Limited explanation about how to enhance access control solutions. | I |
| Disterer et al, 2013 | Opportunities and risks of BYO and comparison of desktop virtualisation models. | Only discusses desktop virtualisation models, with a mere mention of MDM. | I |
| Eslahi et al, 2013 | Discusses MDM, MIM, MAM and Mobile bot nets. | Limitations of MDM, MIM and MAM are not mentioned. | S/R |
| Hansman, 2004 | Taxonomy theories for network security vulnerabilities. | Focus is only on categorising attacks and threats. | I |
| Hormazd, 2014 | Explanation of access control methods that protect data from some threats and attack types. | Advice only revolves around access control initiatives. | I |
| Leavitt, 2013 | Explains mobile specific security frameworks, cloud storage and vulnerabilities. | Only acknowledges a few threats and MDM related end point security methods. | I |
| Morrow, 2012 | Mobile device vulnerabilities as challenges, supported by statistical evidence. | Information is influenced by statistics, thus is biased by trends reported 3 years ago. | I |
| Scarfo, 2012 | Presentation of trends and security frameworks currently favoured by businesses. | Biased towards desktop virtualisation, in comparison to other solutions presented. | I |
| Tokoyoshi, 2012 | Explores issues influencing BYOD policies and ideas for mitigating risks. | Security frameworks are mentioned, yet are not explained in detail. | I |
| Wang et al, 2014 | Specific security frameworks and challenges are discussed. | Frameworks are limited to VPNs and MDM variations. | S/R |

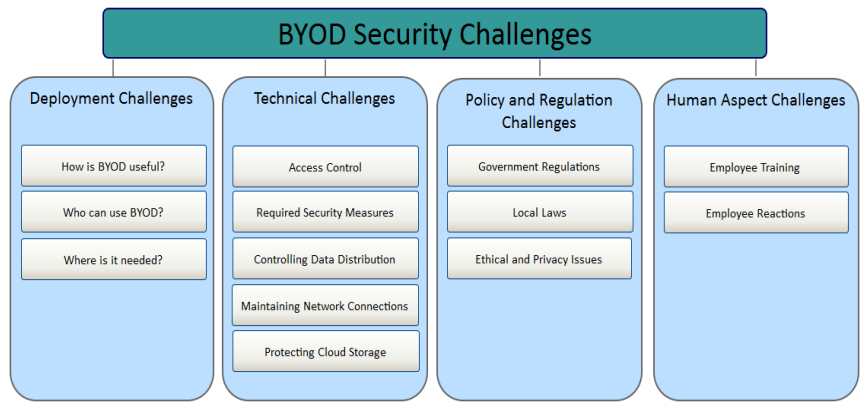

**Figure 1.** Categories of Security Challenges.

**Table 2.** Literature review index based on security challenge focus

| Category | Security Challenge | Research |
|---|---|---|
| Deployment Challenges | Determining how to implement BYOD security measures into existing networks. | [2][7] |
| | Determining who in the organization needs BYOD. | [2][7] |
| | Determining where BYOD is useful. | [2][7] |
| Technical | Access Control. | [2] |

| Challenges | Implementing security measures to protect all device hardware and operating systems. | [8] |
| --- | --- | --- |
| | Providing ongoing support 24/7 | [2] |
| | Containing, controlling, monitoring data distribution. | [22][32][13][28] |
| | Maintaining secure and stable connections. | [22][2] |
| | Protecting cloud storage facilities. | [32][34][25][3][33][36] |
| Policy & Regulation Challenges | Local government regulations and laws. | [1][4] |
| | Ethical and privacy issues. | [1][13] |
| Human Aspect Challenges | Employee training and ongoing education of BYOD security. | [7][17] |
| | Employee reactions, emotions and compliance of BYOD policies. | [8][37] |

**Providing ongoing support** for BYOD security policies demands extra resources to maintain the desired level of security for all devices connected to the network. The responsibilities of security personnel will increase to meet these needs. BYOD security solutions require commitment, time and money, especially during deployment [2].

**Containing, monitoring and controlling the distribution** of data is a primary concern for companies enabling BYOD initiatives [22]. Maintaining confidentiality and integrity of data depends on whether it is stored or only accessed by mobile devices. Monitoring data on devices is complicated as the company loses sight of it once it is transferred from their network, which leads to the potential of data leakage [32].

**Maintaining secure and stable connections** between corporate network resources and devices connected via external networks is a common concern for BYOD reliant businesses. Wireless access points may contain threats such as malware which installs itself on a device when a connection is initiated [22][2]. Factors influencing connections include employees use of public, unprotected Wi-Fi hot spots, and unknown security configurations of their home networks.

**Protecting company data stored on a cloud facility** is a sensitive issue, as cloud applications enable data to be accessed anytime, and may be used as an alternative or eliminates the need to store data on mobile devices [3][33][36]. When cloud based storage is accessed from mobile devices, it is also prone to the same security threats as the device [32], such as hacking, software based attacks, and can also exacerbate other BYOD security challenges such as containing, controlling, monitoring distribution and contamination of data. The inability for a company to control the transfer of data creates security loop holes (cloud sprawl), which occurs when employees transfer company data to public clouds for file sharing, then do not delete later. Cloud service providers also maintain backups of data for reliability reasons, thus data is never completely destroyed [34][20]. The likelihood of threats against cloud storage and mobile devices is increased by enabling the "remember password" feature (storing login credentials in the authentication cache) [34][25].

*C. Policy and Regulation Challenges*

**Local government regulations and laws** regarding corporate data determines rules incorporated into a company's BYOD security policy [1]. Legislations may limit levels of control that companies can enforce on employee owned devices. Companies spread globally need to adjust BYOD policies for each country they are based, in order to align with local laws, which makes streamlining employee contracts and monitoring changing laws more laborious. Legislations influencing BYOD initiatives in Australia include the Privacy Act (1988) and the Freedom of Information Act (1982) [4].

**Ethical and privacy issues** coincide with aforementioned legal implications. When employees provide devices for work use, companies must consider how evasive security measures are, and how they comply with data privacy rights and regulations. Sensitive data needs to be under tight surveillance in order to avoid data leaks which lead to lawsuits [1]. Most data privacy laws worldwide state that employees must provide consent before companies install invasive security measures or access data on personal devices, and in return, the company needs to provide adequate protection [1]. Ideally, security solutions are always active; however it can restrict how employees interact with devices outside of work [13].

*D. Human Aspects Challenges*

**Training and educating employees** about BYOD security, deployed solutions, and enforcement of security policies is critical. This challenge is enhanced when all staff need to have the same understanding of companywide BYOD security policies, yet those handling more sensitive data have extra procedures to follow [7]. Effectively teaching staff in a way they will understand and follow BYOD policies is an ongoing issue. The main aim of training is to convey expectations of acceptable device use, ensure awareness of risks, and how to maintain good security practises [17].

**Employee reactions, emotions and observance of BYOD security policies** is an ongoing challenge for businesses to monitor, contain and maintain [8]. Policies need to include guidelines for handling situations where employees show resistance, utilize mobile devices for illegal activities, or experience difficulty adjusting to them. Over time employees have a tendency to forget guidelines set by policies, or are unaware of changes, which highlights the need for constant reinforcement and training. Employees who strongly disagree with limitations enforced by BYOD security policies, will actively seek loopholes to exploit [37].

**Table 3.** Literature review index based on security framework and solutions focus

| Category | Framework/ Solution Explored | Research |
|---|---|---|
| Comprehensive BYOD Security Frameworks | Company's existing security measures | [9][31][11][39] |
| | Network Access Control (NAC) | [9][12][24][30] |
| | MDM | [23][21][35][11][25] |
| | MAM | [35][25] |
| | MIM | [35][14] |
| | Desktop virtualization models | [35] |
| Single Purpose BYOD Security Solutions | End user agreements, acceptable usage policies, liability agreements | [37][1][2][17][7] |
| | Containerization | [31][20][25][16] |
| | Remote wiping | [25][15] |
| | Anti-malware, anti-virus, anti-spyware solutions | [20] |

### III. EXISTING FRAMEWORKS FOR BYOD SECURITY

Security solutions generally maintain a single focus and are recommended to enhance comprehensive frameworks, which are multi-functional approaches.

*A. Comprehensive BYOD Security Frameworks*

**Existing security measures** include Virtual Private Networks (VPNs), firewalls and email filtering [9][31], are ideal for protecting resources inside networks and when mobile devices are already engaged in BYOD prior to the enforcement of formal policies. VPNs facilitate exclusive network connections with devices and allows access to resources in a controlled environment [11][39]. This reduces the need for storing data on devices, whilst accommodating flexible work patterns. Firewalls protect networks by monitoring network traffic and denying access to suspicious requests. Email filtering detects and warns users of infected emails. Mobile devices can sync email applications, therefore benefiting the device when email filtering is active [31].

**Network Access Control (NAC)** limits the number of connected devices, determines permissions and denies unrecognized devices access to a company's internal network [9][12][24][30]. It was well established prior to the rise of BYOD, yet is pivotal for enhancing BYOD frameworks. Identity and Access Management (IAM) is a variation of NAC, which applies customized device access control rules to a network, yet also manages single sign on and separation of duties [9]. Similarly, Access Application control (AAC), is installed on a mobile device and performs identical access control functions. Desktop virtualization and MDM variations are heavily reliant on NAC and ACC. NAC helps ensure that the probability of data leakage, company-wide malware infections and other attacks are reduced or avoided.

**Mobile Device Management (MDM)** is a multi-functional framework which grants businesses the ability to strictly control mobile devices. MDM solutions contain a main component which manages protocols, provides constant control and monitoring, resides within the company's network and relies on the exchange of certificates to authenticate and communicate with MDM agents, which are installed on mobile devices [23][21]. Together they enforce access rights, update, synchronise files, trigger remote wiping, support VPN connections, conduct anti-malware scans, and provide activity reports [35][11][25]. MDM is useful for companies implementing BYOD strategies rapidly and require a centralized, simplified solution.

**Mobile Application Management (MAM)** is a flexible alternative to MDM, as the scope of protection concerns a specific set of applications on the mobile device. MAM allows the company to apply security policies, lock down, define access control rules, configure software behaviours, remote wipe applications under its control, restrict access to unauthorised applications and install approved applications. Applications outside of MAM's boundaries remain private and continue to function at the employee's discretion [35][25]. MAM is enhanced when combined with containerisation.

**Mobile Information Management (MIM)** is primarily concerned with data integrity and encryption, determines application and personnel access and ensures document synchronization amongst multiple devices, whilst simultaneously administering security procedures such as malware scanning [35]. Company data is located in one place, such as a cloud server, yet is accessed according to permission rules applied to the requesting devices and applications [14]. MIM synchronises data across devices similarly to cloud storage services; as data is stored in a virtual central location.

**Desktop virtualization models** enable desktop computers, virtual machines and servers to host sessions for remotely located devices. Mobile devices operate like remote controls when interacting with applications contained on hosting hardware, and communicate via VPN connections. There are four types of end user virtualisation: virtual desktop streaming, application streaming, hosted virtual desktop & hosted virtual applications, which divide applications, operating systems and user profiles into independent, yet cohesive layers which adapt to user profiles. Some businesses already use this technique, however it is becoming increasingly prevalent as BYOD grows [35]. Desktop virtualisation models are low cost, centralise resources, data and security management and reduces or eliminates the need to transmit data onto mobile devices, thus reduces the possibility of data leakage occurring.

## B. Single Purpose BYOD Security Solutions

**End user agreements, acceptable usage policies and liability agreements** are formal contracts ensuring companies and employees mutually agree upon BYOD security policies; this is vital to the success of BYOD [37]. Agreements support all security controls in place, as they make certain employees know what is expected whilst using personal devices for work, and protects the business on legal accounts in the case of a security breach [1]. BYOD policies contain information such lists of permitted applications, installed security measures, management access, levels of access control, back up procedures, and rules concerning storage of data [2][17]. For example, employees may use VoIP applications, yet social websites are prohibited during work hours. Businesses are advised to involve employees when devising BYOD security policies in order to help them understand responsibilities [7].

**Containerization** partitions mobile device storage space into independent sections in order to divide personal and work data. The section containing company data has its own security policies applied and allows remote access for company control, without affecting personal data [31][20]. The company can also specify a browser within the container to help secure online traffic [25]. Gessner et al. suggests using containerisation as perimeter defence, where its internal applications utilise VPN connections to access resources in the company's network, whilst allowing policy management to direct control. Policy management includes rules controlling access rights of devices, and security procedures required to ensure the contents of the container are protected from threats which may be present elsewhere on the device [16].

**Remote wiping** is the final reactive solution that is triggered when a device is lost, stolen or the owner separates from the company. The technique involves logging into, then removing all company applications and data residing on the device [25][15]. Some commercially available MDM and MAM solutions already contain remote wiping procedures.

**Antivirus, anti-malware and spyware** applications are essential for strengthening BYOD security frameworks [20]. It is imperative that companies enforce the use of these measures and employees using mobile devices for work reasons have some form of this software installed and actively scanning, in order to reduce the chances of infecting resources and other devices connected to the company's network.

## IV. LIMITATIONS OF EXISTING FRAMEWORKS

### A. Limits of Comprehensive BYOD Security Frameworks

VPNs, firewalls and email filtering are biased towards protecting internal network resources. Mobile devices are not fully protected and are still capable of transmitting malware into the network and opening loop holes for other threats such as data leakage. Firewalls and antivirus software installed on company networks may only recognize threats targeting PC operating systems, thus allowing mobile OS specific malware to enter the network and infect other devices. Email filtering is restricted by its dependency on commitment of end users [31]. The primary purpose of NAC is to protect network entry points, and as such cannot single-handedly detect suspicious activity [24]. Once infected applications enter the network, NAC holds little control over its activities [12]. Other down falls include limits on the number of connected devices that can be supported simultaneously and increased strain on administrators who monitor network traffic. Application Access Control is prone to being dismissed by employees as it is an intrusive form of access control [30].

MDM is controversial as all applications and data on the device (work and personal) are subject to security protocols it enforces. It is an endpoint, access control solution whose security features are primarily reactive measures. Lack of preventative measures still leaves mobile devices prone to inappropriate use if stolen or lost [20][25]. Employees are usually resistant of MDM, as they feel restricted, their privacy is invaded or that device ownership is surrendered [25][24]. MDM can be laborious to maintain, as connected devices constantly vary and it requires regular updating [20].

MIM and MAM have similar limitations to MDM in regards to access control and heavy focus towards reactive security measures. Neither offer control like MDM, which limits the company's power to control devices. MAM only protects applications, whilst MIM protects data stored in a central location, and both provide minimum protection against malware [14]. MAM does not explicitly protect data and the placement of its boundaries around selected applications can inhibit communication with personal applications [16]. Businesses must consider management of data synchronisation as employees overriding each others work is a potential consequence if MIM policies are not refined.

Desktop virtualisation models fundamentally depend on stable and secure network connections, and the strength of these affect the safety of transmitted data. If too many employees connect simultaneously to a particular virtual desktop environment, bottlenecks can occur at it's entry points. Businesses still need to determine user access permissions and apply monitoring techniques, such as session management. Network security solutions offer little protection from data leakage and may increase hardware costs.

### B. Limits of Single Purpose BYOD Security Solutions

End user, liability agreements and acceptable usage policies are limited by how strongly the administering company enforces them. Human error, general negligence and failure to comply with BYOD security agreements contribute to risks and damages incurred as a result of security breaches and lost intellectual property [29][37][38][6][8]. Compliance, auditing and agreements that are not BYOD specific are prone to being challenged by resistant employees who disagree or have malicious intent [37].

Containerisation only places boundaries around selected applications and does not prevent employees from copying data in the container to other storage spaces, which means there is no protection from suspicious activities [31]. Remote wiping is only a reactive measure which does not prevent data from being stolen or used for malicious reasons. Remote wiping is obsolete if its execution is delayed; if not triggered immediately after an event, data may already be compromised. Antivirus, antimalware and spyware are reactive measures concerned with counteracting software based attacks as they appear to devices. Their effectiveness is dependent on the device owner's initiative to execute scans regularly. They may not protect the device entirely, due to the rapid rate at which malware is growing, and multiple anti-malware applications may be required, which is time consuming and tedious.

## V. Conclusion

In light of challenges and frameworks discussed, it is evident that BYOD security requires further research and development. Although frameworks discussed are effective, there is room to improve, reduce limitations and close existing loopholes. Scholars recommend implementing a multi layered approach when devising BYOD security policies [31][35][10], yet seldom provide thorough advice about uniting existing frameworks and solutions effectively. It is fair to state that industry awareness needs to gain a higher priority. Existing frameworks will eventually extend themselves to flexibly suit specific business needs in response to cybercrime and the growth rate of malware targeting mobile operating systems.

## References

[1] Absalom, R. (2012) International Data Privacy Legislation Review: A guide for BYOD policies. Ovum. Vol. 1. pp. 1-23.

[2] Astani, M., Ready, K., Tessema, M. (2013) BYOD Issues and Strategies in Organisations. Issues in Information Systems. Vol: 14, Issue 2. pp. 195-201.

[3] Amoroso, EG. (2013). From the Enterprise Perimeter to a Mobility-Enabled Secure Cloud. *Security & Privacy, IEEE*. Vol 1. Pp. 23 - 31.

[4] Australian Government, Department of defence: intelligence and security. (2014) Bring Your Own Device (BYOD) For Executives. Paper explaining guidelines for corporate BYOD policies, submitted online, February 2014, Australia. Pp. 1-3.

[5] Barker, J (2014) Kensington Survey: Majority of organizations report BYOD creates greater security risks. Close-Up media Inc, Coventry, USA, November 2014. pp.1-2.

[6] Beaver K (2012) The BYOD Security Loophole. In *Security Technology Executive*. Vol. May 2012, pp.20.

[7] Bradford Networks (2012) Ten Steps to Secure BYOD. Whitepaper by Bradford Networks, MA, USA, 2012. pp. 1-4.

[8] Chen, H., Li, Hoang, T., Lou, X. (2013) Security challenges of BYOD: a security education, training and awareness perspective. The University of Melbourne, Australia. Pp. 1-8.

[9] Dell Inc (2015) Dell Offers Top Five Best practices for Overcoming BYOD and Mobile Security Challenges. Paper presented to ENP Newswire Publishing, UK. pp. 1-3.

[10] Denman, S. (2012). Why multi-layered security is still the best defence. *Network Security*, Vol 2012. Issue 3. Pp. 5–7.

[11] Disterer G and Kleiner C (2013) BYOD Bring Your Own Device. *Procedia Technology* Vol. 9, 43-53.

[12] Dongwan, K., Changmin, J., Taeeum, K., Hwankuk, K. (2015) A Study on Security framework for BYOD environment. Institute of Research Engineers and Doctors, USA. Pp. 89-92.

[13] Eschelbeck G and Schwartzberg D (2012) BYOD Risks and Rewards: How to keep employee smartphones, laptops and tablets secure. Whitepaper by Sophos, Oxford, UK, June 2012. pp. 1-7.

[14] Eslahi, M., Naseri, M., Hashim, H., Tahir, NM., Mat Saad, E. (2013) BYOD: Current State and Security Challenges. Universitii Teknologi MARA, Malaysia Pp. 1-4.

[15] French A, Guo C and Shim JP (2013) Current Status, Issues, and Future of Bring Your Own Device (BYOD). *Communications of the Association for Information Systems*. Vol. 35. pp. 191-197.

[16] Gessner D, Girao J, Karame G and Li W (2013) Towards a User-Friendly Security-Enhancing BYOD Solution. *Technical Researches, NEC Technical Journal*. Vol. 7. pp. 113-116.

[17] Gladyng, C. (2013) BYOD: Can it harm your business?: A mobile device based study. University of Derby, UK. Pp. 31-34.

[18] Hansman, S., Hunt, R. (2004). *A taxonomy of network and computer attacks*. Computers and Security. Vol. 24. Issue 1. pp. 31-43.

[19] Hoffman, R (2013) Close the BYOD Security Hole. Information Week, United Business Media LLC. Vol. pp. 16.

[20] Hormazd Romer A (2014) Best practices for BYOD security. *Computer Fraud and Security*. Vol. January 2014. pp. 13-15.

[21] Keunwoo, R., Woongryul, J., Dongho, W (2012) Security requirements of a Mobile Device Management System. *International Journal of Security and its Applications*. Vol. 6. pp. 1-6.

[22] Kim K and Hong S (2013) Study on Enhancing Vulnerability Evaluations for BYOD Security. *International Journal of Security and Its Applications*. Vol.8. pp. 299-238.

[23] Kim, S and Jin H (2015) A Simple Security Architecture for Mobile Office. *International Journal of Security and its Applications*. Vol. 9. pp. 139-146.

[24] Koh, E., Oh, J., Im, C. (2014) A study on security threats and dynamic access control technology for BYOD, Smart-work Environment.

[25] Leavitt N (2013) Today's Mobile Security Requires a New Approach. *Technology News, Computer*. Vol. pp. 16-19, IEEE Computer Society.

[26] Lennon RG (2013) Changing User Attitudes to Security in Bring your Own Device (BYOD) and the Cloud. Paper submitted to Computing Department of Letterkenny Institute of Technology, Co Donegal, Ireland. pp.1-4.

[27] Malloy M (2014) Webroot Rolls out New BYOD Security Report. *Wireless News, Close-Up Media Inc*, USA. pp. 1-2.

[28] Mitrovic, Z., Veljkovic, I., Whyte, G., Thompson, K. (2014) Introducing BYOD in an organisation: the risk and customer services view points. The 1st Namibia Customer Service Awards & Conference, November, 2014. pp. 1-26.

[29] Morrow B (2012) BYOD security challenges: control and protect your most sensitive data. *Network Security*. Vol. December 2012. pp. 5-8.

[30] Pell, L. (2013) BYOD Implementing the Right Policy. University of Derby, UK. Pp. 95-98.

[31] Rhodes J (2013) Building Security Around BYOD. *Managing Mobility, Rough Notes*. Vol. 156. pp. 104, 114.

[32] Rodríguez, NR., Murazzo, MA., Chavez, S. (2012). Key aspects for the development of applications for Mobile Cloud Computing. *Journal of Computer Science & Technology*. vol. 13, no. 3. pp. 143-148.

[33] Sahu, D., Sharma, S., Dubey, V., Tripathi, A. (2012). Cloud Computing in Mobile Applications. *International Journal of Scientific and Research Publications*, Vol 2. Issue 8. August 2012. pp. 1-9.

[34] Samaras V, Daskapan S, Ahmad R and Ray S (2014) An Enterprise Security Architecture for Accessing SaaS Cloud Services with BYOD. Paper submitted to Delft University of Technology, Netherlands and Manukau Institute of Technology, New Zealand. pp. 1-6.

[35] Scarfo A (2012) New Security perspectives around BYOD. *2012 Seventh International Conference on Broadband, Wireless computing, Communication and Applications*. Vol. pp. 446- 451, IEEE Press.

[36] Subramanian, L., Maguire Jr, GQ. (2011). An architecture to provide cloud based security services for smartphones. 27th Meeting of the Wireless World Research Forum (WWRF), *Wireless World Research Forum, 2011Conference paper* (Refereed).

[37] Thomson G (2012) BYOD: enabling the chaos. *Network Security*. Vol. February 2012. pp. 5-8.

[38] Tokuyoshi B (2012) The security implications of BYOD. *Network Security*. Vol. April 2013. pp. 12-13.

[39] Wang W, Wei J and Vangury K (2014) Bring Your Own Device Security Issues and Challenges. Paper presented to The 11th Annual IEEE CCNC- Mobile Device, Platform and Communication, USA. pp. 80-85.